\documentclass[
reprint,
superscriptaddress, 
showpacs,
amsmath,amssymb,
aps,
prl,
]{revtex4-2}
\newcommand{\id}{\mathds{1}}
\newcommand{\mG}{\mathcal{G}}
\newcommand{\mH}{\mathcal{H}}
\newcommand{\mV}{\mathcal{V}}

\usepackage{booktabs}
\usepackage{amsmath,amsthm,dsfont,graphicx}
\newtheorem{lemma}{Lemma}
\newtheorem{definition}{Definition}
\newtheorem{theorem}{Theorem}
\usepackage[caption=false]{subfig}
\usepackage{amssymb,amsfonts}
\usepackage{mathrsfs}
\usepackage{graphicx,xcolor,multirow}
\usepackage{dcolumn}
\usepackage{appendix}
\usepackage{verbatim}
\usepackage{bm}

\makeatletter
\newcommand{\tr}{{\rm Tr}}
\newcommand{\ty}[1]{{\lceil{#1}\rceil}}
\renewcommand{\maketag@@@}[1]{\hbox{\m@th\normalsize\normalfont#1}}%

\makeatother

\begin{document}


\title{Exploring the boundary of quantum network states from inside out}

\author{Xiang Zhou}
\affiliation{Hefei National Laboratory for Physical Sciences at Microscale and Department of Modern Physics, University of Science and Technology of China, Hefei Anhui 230026, China}
\author{Zhen-Peng Xu}\email{zhen-peng.xu@ahu.edu.cn}
\affiliation{School of Physics and Optoelectronics Engineering, Anhui University, Hefei 230601, China}
\author{Liang-Liang Sun}
\affiliation{Hefei National Laboratory for Physical Sciences at Microscale and Department of Modern Physics, University of Science and Technology of China, Hefei Anhui 230026, China}
\author{Chunfeng Wu}
\email{chunfeng\_wu@sutd.edu.sg}
\affiliation{Science, Mathematics and Technology, Singapore University of Technology and Design, 8 Somapah Road, Singapore 487372, Singapore}
\author{Sixia Yu}
\email{yusixia@ustc.edu.cn}
\affiliation{Hefei National Laboratory for Physical Sciences at Microscale and Department of Modern Physics, University of Science and Technology of China, Hefei Anhui 230026, China}
\affiliation{Hefei National Laboratory, University of Science and Technology of
China, Hefei 230088, China}
\date{\today}

\begin{abstract}
Quantum networks with bipartite resources and shared randomness present the simplest infrastructure for implementing a future quantum internet. Here, we shall investigate which kinds of entanglement can or cannot be generated from this kind of quantum network by examining their fidelity with different graph states. On the one hand, based on a standard form of graph states under local complementation and a fine-grained uncertainty relation between two projections, we establish upper bounds of fidelity that improve over previous results by at least 25\% as the dimension of local systems tends to infinity. On the other hand, in the triangle network, we propose efficient protocols to generate genuine multipartite entangled states from the network, providing significant nontrivial lower bounds of fidelity with high dimensional GHZ states.

\end{abstract}

                              
\maketitle

\textit{Introduction.---}
The second quantum revolution~\cite{dowling2003quantum,jaeger2018second,aspect2023second} ushered by strange and extraordinary quantum phenomena like quantum entanglement~\cite{horodecki2009quantum} and nonlocality~\cite{brunner2014bell} is spawning new technologies. The outstanding representatives are quantum cryptography~\cite{gisin2002quantum}, quantum computing~\cite{ladd2010quantum}, quantum communications~\cite{gisin2007quantum} and quantum metrology~\cite{giovannetti2011advances}. 
Quantum internet~\cite{kimble2008quantum} has also been envisioned as a platform armed by all those quantum technologies.

Lasting theoretical effort on quantum networks has been devoted in the last decades,
including the studies of quantum repeaters~\cite{briegel1998quantum}, quantum conference key agreement~\cite{murta2020quantum}. Very recently, the network version of quantum correlations~\cite{navascues2020genuine,aberg2020semidefinite,kraft2021quantum,renou2019genuine,contreras2021genuine,tavakoli2021bell,mao2022test,jones2021network,xu2023characterizing,kraft2021characterizing}
as entanglement and nonlocality are attracting further attention, which deepens the understanding of quantum foudations~\cite{renou2021quantum} and is beneficial for applications~\cite{vsupic2023quantum}.
A distributed network is used in the setup of network correlations, represented by a hypergraph with hyperedges and nodes standing for sources and receivers, respectively, and a source only sends particles to the receivers if the corresponding nodes are contained in the corresponding hyperedge. For a network with just bipartite sources, the usage of hypergraph reduces to normal graph. 
Only shared randomness but no classical communications among nodes are allowed in such setup.
The reasons in practice can be twofold. On the one hand, classical communications in a long-distance network require reliable quantum memories~\cite{brown2016quantum}, which is still a challenge nowadays. On the other hand, private communications are not permitted in some tasks, and all the distribution of resources is controlled by a center.
The quantum states preparable in this way with quantum sources are \textit{quantum network states}, while the unpreparable ones are \textit{quantum network entangled}~\cite{navascues2020genuine}. 

The detection of quantum network entanglement has been at the core of this direction since its beginning. Various methods emerge for this purpose, including semidefinite programming based on inflation~\cite{navascues2020genuine}, covariance decomposition~\cite{jones2021network,kraft2021characterizing,xu2023characterizing}, uncertainty relations~\cite{hansenne2022symmetries,wang2024quantum,makuta2023no} and inequalities of network locality~\cite{mao2022test}.
As it turns out, GHZ states, graph states with prime local dimensions, entangled symmetric states etc. are quantum network entangled~\cite{navascues2020genuine,hansenne2022symmetries}. Witness of quantum network entanglement based on the fidelity with GHZ states and general graph states has also then been proposed and improved~\cite{makuta2023no,wang2024quantum}. 
It worths to mention that the appliacation of quantum network states in quantum metrology is also limited~\cite{yang2024quantum}. 

All those results characterize the set of quantum network states from outside. In comparison, few results exist to describe such a set from the inside.
Only protocols with local operations limited to unitaries exist~\cite{kraft2021quantum},
with which the prepared
quantum network states has fidelity with GHZ states in dimensions $2$ to be $1/2$. However, such a fidelity can even be $0.51704$ when the local operations are general channels as shown without explicit protocol~\cite{navascues2020genuine}. Besides further improvement on the detection of quantum network entanglement, more research from the inner side, especially, explicit protocols with general channels are still missing. 

In this work, we first establish a standard form of high-dimensional graph states based on the equivalence from local complements. Then, by developing and applying techniques in fine-grained uncertainty relations, we substantially improve the upper bounds of fidelity between arbitrary graph states and network states with bipartite sources. Finally, we propose explicit protocols to generate quantum network states with genuine multipartite entanglement.

\textit{Detecting network entanglement via graph states.---}
A general distributed quantum network can be represented by a hyergraph, where each node stands for a player which applies a local operations on all the incoming particles, and each hyperedge stands for a quantum source which distributes particles to players corresponding to nodes in that hyperedge. Besides, a global classical random variable is allowed, interpreted as a shared protocol among players in advance and then no classical communication is possible between players. This happens when success classical communication consumes too much resources for the players far away from each other, or the time window for classical communication is too short like in high frequency local operations. There could be even a global controller who forbids any classical communication among players. 

Through the whole paper, we only consider distributed quantum networks with bipartite sources, in which case the associated hypergraph is just a normal graph and hyperedges are normal edge with two nodes. Without loss of generality, the consideration can reduce to quantum networks with complete graphs, since any other quantum networks with bipartite sources can be regarded as a special case by tuning some sources to send no effective particles.
A general $n$-partite state prepared in such a quantum network has the form:
\begin{equation}\label{2a2}
    \rho = \sum\nolimits_{\lambda}p_{\lambda} \left(\otimes_{i=1}^n \mathcal{E}_i^\lambda\right)[\otimes_{j>i}\sigma_{ij}],
\end{equation}
where $\lambda$ is the shared classial random variable, $\mathcal{E}_i^\lambda$ is the channel of $i$-th player and $\sigma_{ij}$ is the source distributed between players $i$ and $j$. We say any state which cannot be written as in Eq.~\eqref{2a2} to be network entangled for bipartite sources (NEBS).
We remark that any entangled state can be prepared in the quantum networks with bipartite sources once classical communications are allowed. The amount of classical communications can even provide measures of network entanglement~\cite{xu2024quantum}. More economic protocols with less classical communications is another research direction. Here we focus on the fidelity between high-dimensional graph states and states from quantum network without classical communications.

 As it turns out, many interesting states including GHZ states, qubit graph states and entangled symmetric states are NEBS. Any high-dimensional graph states with local dimension to be a prime number are also NEBS, whose fidelity with network states in Eq.~\eqref{2a2} has also been upper bounded~\cite{wang2024quantum,makuta2023no}.
A high-dimensional graph state can be defined via a multigraph $G = (V, \Gamma)$ with a vertex set $V$ and an adjacency matrix $\Gamma$ specifying how vertices are connected. Each entry $\Gamma_{ij}$ of the adjacency matrix denotes the number of edges connecting vertices $i$ and $j$ for $i\not=j$ and $\Gamma_{jj}=0$ as there are no self-loops. Let $N_i$ be the neighborhood of vertex $i$, i.e., a subset of $V$ of all vertices connecting to $i$ with a positive number of edges. An example of 3-vertex multigraphs is shown in Fig.3(a) for which we have a double edge $\Gamma_{12}=2$ and a normal edge  $\Gamma_{13}=1$ with the neighborhood of vertex 1 being $N_1 = \{2, 3\}$.

\begin{figure}
    \centering
    \includegraphics[width=0.43\textwidth]{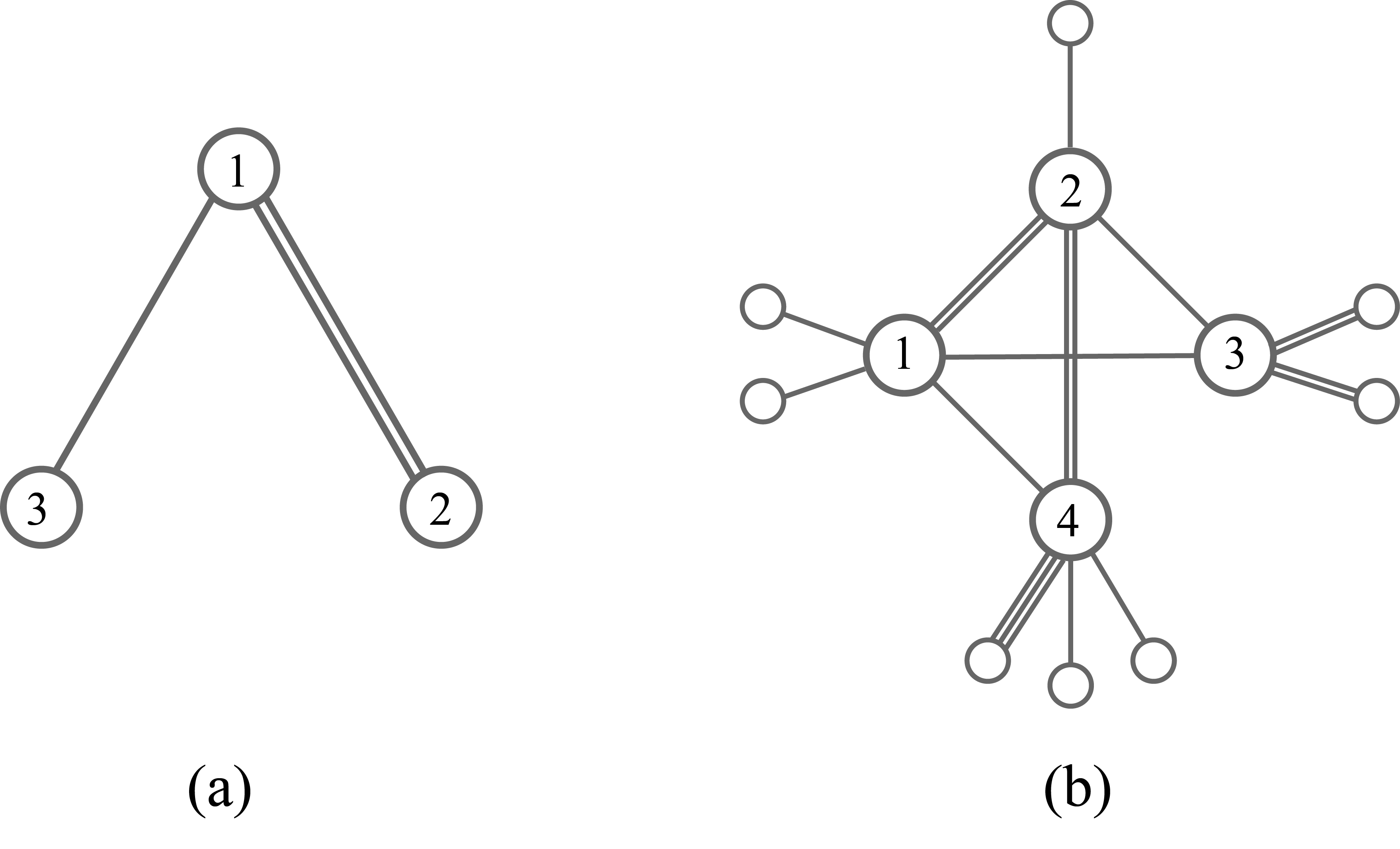}
    \caption{Two connected multigraphs representing two qudit graph states with different indices (a) $\beta=1$ and (b) $\beta=5$, where $\beta=2|N_1\cap N_2|+1$ and $N_i$ denotes the neighborhood of $i$. }
    \label{fig:thm1}
\end{figure}

Let each vertex in $V$ label a qudit with a prime dimension $d$, and for each qudit, we denote by $X=\sum_{i=0}^{d-1}|i+1\rangle\langle i|$ and $Z=\sum_{i=0}^{d-1}\omega^i|i\rangle\langle i|$ with $\omega=\exp(2\pi i/d)$ the generalized Pauli matrices.
By assigning to each vertex $i\in V$ a string of generalized Pauli operators 
\begin{equation}\label{2c}
    g_i=X_i\bigotimes_{j\in V}Z_i^{\Gamma_{ij}},
\end{equation}
where the subscripts of generalized Pauli operators specify the qudits they acting on, we can define the graph state $|G\rangle$ associated to the multigraph $G=(V,\Gamma)$ to be the unique state that is stabilized by all these vertex operators, i.e., $g_i|G\rangle=|G\rangle$ for all $i\in V$ \cite{hein2006entanglement}.

From one multigraph, one can obtain another multigraph by performing an operation called local complementation (LC), after which the corresponding graph states of two multigraphs are equivalent under local unitary transformations~\cite{keet2010quantum}. In particular, an LC around vertex $l\in V$ with a weight $a\in F_d=\{0,1,\ldots,d-1\}$  sends a multigraph $G=(V,\Gamma)$ to the multigraph $G'=(V,\Gamma')$ with its adjacency matrix $\Gamma'$ given by entries
$$\Gamma_{ij}'= \Gamma_{ij}+a\Gamma_{il}\Gamma_{jl},\quad (i\not=j),$$
where the additions and multiplications are operated in the field $F_d$.  One can perform the same local complement multiple times on the same vertex or on different vertices in any order. Our first result is about a standard form of multigraphs and corresponding graph states.

\begin{theorem}\label{thm:class}
Under local complementations,  all connected graphs with no less than three vertices can be transformed to a standard form in which there is a pair of connected vertices 1 and 2 such that every vertex $u\in \{1\}\cup N_1\cap N_2$ has a neighbor different and disconnected from vertex 2.
\end{theorem}

For convenience, we shall refer to $\beta=2|N_1\cap N_2|+1$ as the index of the graph (state), and in Fig.~\ref{fig:thm1}, we show two graphs with index $\beta=1$ and $\beta=5$. The basic idea for the proof is that whenever we have a vertex $u\in N_1\cap N_2$ whose neighbors different from vertex 2 are connected to vertex 2, we can always perform an LC around $u$ with a suitable weight such that either one obtains a graph with index 1 or disconnects some common neighbors from 2. The process is finite as the size of neighborhood $N_2$ is strictly decreasing. See the Appendix for detailed proof. 
Hence, for a given dimension $d$, we only need to focus on graph states in their standard form to find out the upper bounds of their fidelity with network states of form in Eq.(\ref{2a2}), which is our next main result.

\begin{theorem}
{\it If the fidelity between a state $\rho$ and an arbitrary connected graph state $|G\rangle$ with index $\beta$  satisfies
\begin{equation}\label{eq:upperbound}
 \langle G|\rho |G\rangle > \frac{1+2\sqrt{2\beta d}+(\beta+1)d}{(\beta+2)d+2\sqrt{2\beta d}}
\end{equation}
then $\rho$ cannot be generated in an LOSR quantum network with bipartite sources. 
}
\end{theorem} 

\begin{figure}
    \flushleft
    \includegraphics[width=0.45\textwidth]{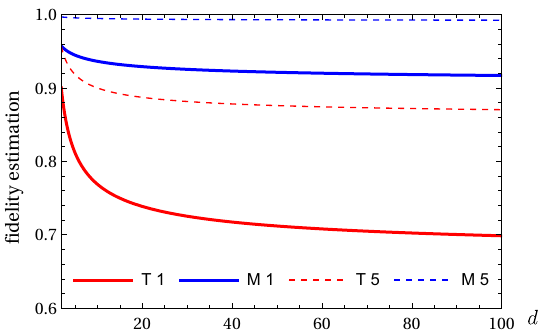} 
    \caption{Comparison of the estimations of the fidelity between multipartite graph states and network states with bipartite sources for a prime local dimension. The two curves in red are ${\rm ub}_2$ from Eq.~\eqref{eq:upperbound}, and the two in blue are ${\rm ub}_1$ from Eq.~\eqref{eq:upperbound1}. The two solid curves correspond to $\beta=1$, and the two dashed correspond to $\beta=5$. }\label{fig:upperbound}
\end{figure}

Our essential tool for the proof is a fine-grained uncertainty relation (FiGUR), in addition to the inflation technique. This type of uncertainty relation involves the complete set of 
probability distributions corresponding to different outcomes of given observables \cite{oppenheim2010uncertainty} and turns out to determine the nonlocality in many scenarios \cite{PhysRevA.109.022408}. For observables with only two outcomes, we have
\begin{lemma}
    For a state $\rho$ and two projectors $P$ and $Q$ satisfying $PQP=\lambda P$ for $0<\lambda<1$, it holds
\begin{equation}\label{eq:fgur}
    \sqrt{1-\tr\rho P}\ge \sqrt{(1-\lambda)\tr\rho Q}-\sqrt{\lambda(1-\tr\rho Q)}.
\end{equation}
\end{lemma} 
To employ this FiUGR for a given graph state for qudits, we take projectors to be the ones corresponding to the $+1$ eigenspace $\lceil g\rceil=\sum_{k=0}^{d-1} g^k/d$ of some operator $g$ satisfying $g^d=1$. For two commuting strings of generalized Pauli operators $g,h$, it holds constraint $\ty{gh}\ge \ty g+\ty h-1$. It turns out that the $+1$ eigenspaces of two noncommuting strings of generalized Pauli operators always satisfy the condition in Lemma 1.
To obtain projectors that do not commute with each other from the stabilizers of a given graph state, a series of inflations of the given network, for which the extra neighbor of a vertex in $\{1\}\cup N_1\cap N_2$ is essential, is employed for the graph states in the standard form. The detailed proof is presented in the Appendix, along with a comparison with a previous classification of graph states~\cite{makuta2023no}.

In Ref.~\cite{makuta2023no}, graph states were classified into four groups, all of which turned out to be able to be transformed into the standard form by LCs, e.g., its third group can be transformed into graph with index 1 (see Appendix). Based on their classification,  
the maximal fidelity is upper bounded by ~\cite{makuta2023no}
\begin{equation}\label{eq:upperbound1}
 \langle G|\rho |G\rangle \le  1-\frac{1}{16}(\sqrt{\beta^2+4\gamma}-\beta)^2
\end{equation}
 where $\gamma=\sqrt{d}/(\sqrt d+1)$ (See Appendix C in Ref.~\cite{makuta2023no} for more details.)  
As illustrated in Fig.~\ref{fig:upperbound}, the upper bound $\mathrm{ub}_2$ defined by the right-hand-side of Eq.(\ref{eq:upperbound}) improves significantly over the upper bound $\mathrm{ub}_1$ defined by the right-hand-side of Eq.(\ref{eq:upperbound1}). For $\beta=1$, ${\rm ub}_2$ goes to $2/3$ as $d$ tends to infinity, with the improvement quantified by $1-\mathrm{ub}_2/\mathrm{ub}_1 = (4-\sqrt{5})/3\sqrt{5} \approx 26.3\%$. 
For a fixed dimension $d$, the distinction becomes more clear as $\beta$ increases, as $(1-{\rm ub}_2)/(1-{\rm ub}_1)$ goes to infinity as $\beta$ tends to infinity. For $\beta=5$, as illustrated in Fig.~\ref{fig:upperbound}, the upper bound ${\rm ub}_1 > 0.99$, which is almost useless for a noisy situation. However, the upper bound ${\rm ub}_2$ is still clearly different from the trivial upper bound $1$. Consequently, the inequality in Eq.~\eqref{eq:upperbound} can detect significantly more NEBS states via their fidelity with graph states. We remark that the result in Eq.~\eqref{eq:upperbound} holds for any high-dimensional graph states with local dimension to be prime.

\textit{Creating entanglement from triangle network.---}
Most effort in the field of quantum network states has been devoted to the characterization from outside, while rare results exist for the characterization from inside~\cite{kraft2021quantum,navascues2020genuine}.
Though the fidelity between $2$-dimensional GHZ state and triangle network state can be $0.51704$~\cite{navascues2020genuine}, explicit protocol for construction of the corresponding network state is still missing.
In this work, we provide efficient protocols to prepare quantum network states of high fidelity with GHZ states in triangular network for different target dimension $d\ge 2$. 

\begin{figure}
  \subfloat[][]{
  \includegraphics[width=0.2\textwidth]{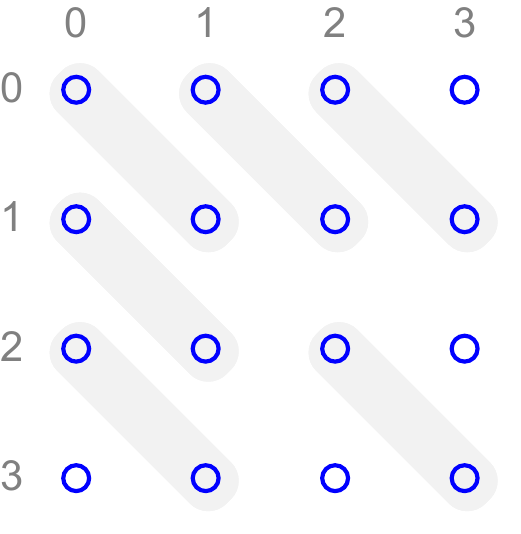}
}\hspace{2em}
\subfloat[][]{
  \includegraphics[width=0.2\textwidth]{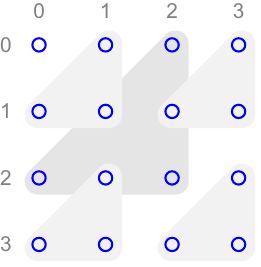}
}
  \caption{The illustration of the measurement part in (a) $d=2$ and (b) $d=3$ protocols when each pair of nodes shares a $4$ dimensional bipartite state. Each block of size one, two, and three corresponds to a projective measurement of dimension one, two, and three, respectively.  In fact, the projectors related to triangles in light gray act on one qubit and the projector related to triangle in dark gray acts on another qubit if we treat each $4$-dimensional particle as a composition of two qubits.}
  \label{fig:protocol4}
\end{figure}

In the case of $d=2$, we let each source generate an entangled pair with local dimension $t$ and send the particles to receivers according to the triangle network topology. Each node receives two particles and carries out a projective measurement $\{\Pi_{a,b}+\Pi_{a+1,b+1}\}_{(a,b)\in M}$ with $\Pi_{ab}=|a,b\rangle\langle a,b|$ and $ M:=\{(a,b)\mid \min\{a,b\} \mbox{ is even}\}$,
where we adopt the convention that $|j\rangle=0$ whenenver $j\ge d$. 
   For the case that $t=4$, the projective measurement is illustrated in Fig.~\ref{fig:protocol4}(a) in which we note that there is only a 1-dimensional projection $\Pi_{03}$ for $(a,b)=(0,3)$ due to the convention above.  After the measurement outputting an outcome $(a,b)$ we apply the following encoding for each node
$$E_{ab}=\sum_{\sigma=0}^{1}|\sigma\rangle\langle a+\sigma,b+\sigma|,\quad (a,b)\in M.$$ 

By optimizting the shared bipartite entangled state, the final fidelity with the $2$-dimensional tripartite GHZ state can be numerically found to be $0.548048$, which goes beyond the known one $0.51704$. Once the fidelity of a given state with a $d$-dimensional GHZ state is higher than $1/d$, there is certainly genuine multipartite entanglement. Though the prepared tripartite states with local dimension $d=2$ does not outperform $1/2$ too much, they can indeed be used to reveal nonlocality and even genuine tripartite nonlocality as detailed in Appendix.

In the case of $d=3$, we suppose that each source prepares $k$ pairs of qubits. For a general local dimension $t$, we can always embed it into the space of $\lceil \log t \rceil$ pairs of qubits. We then apply the following protocol:
\begin{enumerate}
  \item Each source generates $k$ pairs of entangled qubits and send the particles to receivers according to the triangle network topology in order;
  \item Each node receives $2k$ particles in order and group them into $k$ pairs;
  \item Each node carries out the projective measurement $\{\Pi_0 = |00\rangle\langle 00|, \Pi_1 = \id - |00\rangle\langle 00|\}$ on one pair of qubits;
  \item If the outcome is $1$, the $j$-th node maps $|01\rangle, |11\rangle, |10\rangle$ to $|0\oplus i\rangle, |1\oplus i\rangle, |2\oplus i\rangle$, where $x\oplus y$ stands for $(x+y) \mod 3$;
  \item If the outcome is $0$ and it is not the last pair, the node throw away the current pair and continue with step 2-4 with another pair.
  \item Otherwise, we just map $|00\rangle$ to $|0\rangle$. 
\end{enumerate}
The whole procedure of measurements in those two protocols can be regarded as one global projective measurement, which is illustrated in Fig.~\ref{fig:protocol4}. After an optimization over a subset of bipartite sources, the final fidelity turns out to be $2\sqrt{3}-3 \approx 0.464102 > 4/9$. 
For small dimension $t$ of the distributing particles, the corresponding fidelity coincides with the optimal one obtained numerically via a see-saw method.

In the case of $d=k^2$ for some integer $k\ge2$, we suppose that each pair of nodes shares a $k$-dimensional maximally entangled state $|\Psi\rangle = \sum_{i=0}^{k-1} |ii\rangle_{ab}/\sqrt{k}$ with $(a,b)=(1,2),(3,4),(5,6)$. Thus particles $1,6$ are at the first node, particles $2,3$ are at the second, while particles $4,5$ are at the third. Now we apply the following unitary protocol, which is a generalization of the one for $k=2$~\cite{kraft2021quantum},
\begin{equation}~\label{eq:protocol}
    {\rm SWAP}_{16} \otimes {\rm CP}_{23} \otimes {\rm CP}_{54}\ |\Psi\rangle_{12} \otimes |\Psi\rangle_{34} \otimes |\Psi\rangle_{56},
\end{equation}
where the subscript denotes particles. Here ${\rm SWAP}_{16}$ exchanges particles 1 and 6 while ${\rm CP}|ij\rangle_{23} = |i\rangle_2\otimes|j-i\rangle_3$ denotes a controlled permutation with source 2 and target 3 and so on.  The fidelity between $d$-dimensional GHZ state and such a triangle network state is  $1/\sqrt{d}$.

By embedding the state in Eq.~\eqref{eq:protocol} to higher dimensional space, we obtain a lower bound of fidelity for arbitrary dimension $d$, i.e., $\lfloor \sqrt{d} \rfloor/d$. For $d$ smaller but close to $k^2$, another approach is to project the state in Eq.~\eqref{eq:protocol} to dimension $d$, e.g., the fidelity is $(k+2(1-k^2)/k^3)/t$ for $d=k^2 -1$ which is always better than $\lfloor \sqrt{d} \rfloor/d = (k-1)/d$. An even better lower bound can be obtained by replacing state $|\Psi\rangle$ by $\sum_{i=1}^{k-1} |ii\rangle \sqrt{(1-x^2)/k} + x |00\rangle$ and optimizing over $x$. For example, the lower bound is $0.45798$ for $d=3$, which is close to $0.464102$.

\textit{Conclusion.---}
On the one hand, we proposed a new criteria for the states from quantum network with bipartite sources, which has significantly improved the existing upper bound on the fidelity of network states and graph states for any prime local dimension and any number of parties. Consequently, this implies a more powerful tool to detect quantum network entangled state for bipartite sources. Since the upper bound of fidelity is based on a new introduced fine-grained uncertainty relation (FiGUR) and the inflation of high-dimensional graph states in the standard form, more FiGURs and better inflation procedures could be explored further in future.

On the other, we proposed simple protocols to construct quantum triangle network states with genuine multipartite entanglement, aimed at maximal fidelity with the GHZ states for local dimensions $2, 3$ and beyond. With our protocols, the final fidelity for dimension $2$ is $0.548048$ and the one for dimension $3$ is $0.464102$. Such states can even reveal genuine multipartite nonlocality.
The constructions have well-organized patterns as in Fig.~\ref{fig:protocol4}, which leads to an interesting direction to explore in the future, i.e., to figure out similar protocols for other local dimensions, or even unify them in a consistent way.
For the dimension $d\ge 4$, especially when $d$ is lower but close enough to a square number, the proposed projection protocol performs also good enough but still not the best one as we can see for $d=3$, leaving space for future exploration.

\textit{Acknowledgement.---} We thank Nikolai D. Wyderka and Hermann Kampermann for discussion. LLS and SY would like to thank the support from Innovation Program for Quantum Science and Technology (2021ZD0300804). ZPX acknowledges support from National Natural Science Foundation of China (Grant No. 12305007), Anhui Provincial Natural Science Foundation (Grant No. 2308085QA29). { CW is supported by the National Research Foundation, Singapore, and A*STAR under its Quantum Engineering Programme (NRF2021-QEP2-02-P03).}

Note added: By finishing this work, we noticed that Justus Neumann et al. have also worked on the similar topic with  different techniques and related but complementary results.

\appendix{}
\onecolumngrid
\section{A standard from of qudit graph states}

{\it Proof of Theorem 1.---} As the graph is connected and has no less than three vertices, we have either a subtree of length two, say $3-1-2$, or a triangle, i.e., a mutually connected triplet $\{1,2,3\}$. In the latter case, we can always remove the edge between 2 and 3 by an LC around vertex 1 with a suitable weight. Thus, one can assume without loss of generality the existence of a subtree $3-1-2$ such that $1$ is connected to both $2,3$ with $2,3$ disconnected.  For convenience, we call a vertex $u\in \{1\}\cup N_1\cap N_2$ {\it good} if there is a neighbor of $u$ that is different and disconnected from 2 and {\it bad} otherwise. The theorem amounts to prove all vertices in $\Delta_0= N_1\cap N_2$ can be turned into good uncer LCs, as vertex 1 has been chosen above to be a good vertex with 3 being the extra neighbor.
 
 If the neighborhoods of 1 and 2 are disjoint, i.e., $\Delta_0=\emptyset$, then the theorem follows immediately. Otherwise,  we have a nonempty intersection $\Delta_0 = N_1 \cap N_2$ and suppose that there is a bad vertex $u\in \Delta_0$, i.e.,  $N_u\subseteq N_2\cup \{2\}$ all neighbors of $u$ are connected to 2 or vertex 2 itself. {For each vertex $v\in N_u\backslash \{2\}$ the ratio 
 \begin{equation}\label{av}
 a_v=-\frac{\Gamma_{2v}}{\Gamma_{2u}\Gamma_{uv}}
 \end{equation}
is well-defined and nonvanishing. If $a_v$ is independent of $v\in N_u\backslash \{2\}$, then by an LC around $u$ with weight $a=a_v$ we can remove the edges between 2 and all the vertices in the neighborhood $N_u$ of $u$, including vertex 1, since
$$\Gamma_{2j}'=\Gamma_{2j}-a\Gamma_{2u}\Gamma_{ju}$$ is vanishing for all $j=v\in N_u\backslash \{2\}$ in the transformed multigraph. As a result, we have an empty intersection $N_u\cap N_2$, and by taking $1-u-2$ as starting points, we arrive at a graph with index 1. Suppose that there exists $v\in N_u\backslash \{2\}$ such that $a_1\not=a_v$.
An LC around $u$ with weight $a_v$ removes the edge between $2$ and $v$, i.e., $v$ is the neighbor of $u$ disconnected from $2$, while keeping $1,2$ connected. In this way, vertex $u$ turns from bad to good. In the meantime, other good vertices either remain good or are disconnected from vertex 2, i.e., removed from $\Delta_0$. In particular, suppose that $u$ is good with $j_u$ as an extra neighbor disconnected from 2 and after an LC around a bad vertex $v\in \Delta_0$, since by definition $j_u\not\in N_2$ it follows that $j_u\not \in N_v\subseteq N_2$ so that $j_u$ is still disconnected from vertex 2. Note that the good vertex $u$ may be removed from the neighborhood $N_2$ of vertex 2. After this step, the intersection $\Delta_0$ might be changed, i.e., some vertices in $N_2$ disconnected from vertex 1 might be connected with vertex 1 after this LC. However, as all neighbors of a bad vertex $u$ are connected to $2$ or vertex $2$ itself, LCs around $u$ do not increase the neighborhood of $2$. On the contrary, the neighborhood $N_2$ is decreased by at least one vertex, e.g., vertex $v$ is removed from the neighborhood of $2$.

The main step for turning bad vertices into good ones can be summarized as follows. Whenever there is a bad vertex $u\in \Delta_0$, we can either have a graph with index 1 in the case of $a_v$ being constant over $N_u\backslash \{2\}$ or turn this bad vertex $u$ into a good vertex and decrease the neighborhood $N_2$ by at least one vertex by applying an LC around $u$ with a suitable weight. This procedure does not affect those surviving pre-existing good vertices. Thus, after a finite number of steps, all vertices in $\Delta_0$ can be turned into good ones.

{\it Comparison with previous classification ---} According to Ref. \cite{makuta2023no} every graph that has at least three vertices and has at least one node, say vertex 1, with a neighborhood $|N_1|\ge 2$ can be transformed using local complementations and relabelling into a graph $G$ that fulfills $|N_1\backslash N_2|\ge 2$, nodes $1, 2$ are connected and one of the four following sets of conditions:
\begin{description}
\item[${\mathcal{G}}_0$] $N_1\cap N_2=\emptyset$,
\item[${\mathcal{G}}_1$] $N_1\cap N_2\neq\emptyset$ and for all $u\in N_1\cap N_2$ we have $N_2\backslash \{u\}\neq N_u\backslash \{2\}$,
\item[${\mathcal{G}}_2$] $|N_1\cap N_2|=1$ and there exists $u\in N_1\cap N_2$ such that $N_2\backslash \{u\}= N_u\backslash \{2\}$,
\item[${\mathcal{G}}_3$] $|N_1\cap N_2|\ge 2$, there exists $u\in N_1\cap N_2$ such that $N_2\backslash \{u\}= N_u\backslash \{2\}$, and there exists $a\in\{1,...,d-1\}$, depending on $u$, such that for all $v\in N_u\backslash  \{2\}$ we have $\Gamma_{2,v}+a\Gamma_{2,u}\Gamma_{uv}=0$.
\end{description}

As $|N_1\backslash  N_2|\ge 2$ and $2\in N_1$ we have at least a vertex, say vertex 3, that is a neighbor of vertex 1 disconnected from vertex 2 so that we have starting configuration $3-1-2$. \begin{itemize}
\item The graph state in group $\mG_0$ is already standard and has an index of 1.
\item For graphs in group $\mG_1$ we have a good vertex if there exists $v\in N_u\backslash \{2\}$ such that $v\not \in N_2$ and a bad one if $N_u\backslash \{2\}\subset N_2$. In this case, we call for the procedure of turning a bad vertex into a good one as described above to arrive at the standard form.  
\item In group $\mG_2$ we have a single bad vertex $u$ and i) if $a_v$ as defined in Eq.(\ref{av}) is independent of $v\in N_u\backslash \{2\}$ an LC around $u$ with weight $a=a_v$ leads to a graph with index 1; ii) otherwise there exists $v$ such that $a_1\not=a_v$ so that an LC around $v$ with weight $a_v$ removes the edge between $v$ and $2$, making $u$ a good vertex, while keeps $1$ and $2$ connected. 
 At this stage, one might end up with a new intersection $\Delta_0'\subseteq N_2\backslash \{v\}$ and after applying the procedure of turning bad vertices into good ones a finite number of times, one can arrive at a standard form. By treating those bad vertices in graphs belonging to the group ${\cal G}_1$ in the same way, we end up with a standard form. 
\item For graph states in group $\mG_3$, there exists a bad vertex $u$ with $a_v$ independent of $v\in N_u\backslash \{2\}$ so that an LC around vertex $u$ with weight $a=a_v$ removes all edges between vertex $2$ and the neighborhood $N_u$ of vertex $u$, i.e., $N_u'\cap N_2'=\emptyset$. By choosing $1-u-2$ as starting points we have a graph state with index 1. 

\end{itemize}

\section{Proof of Lemmas}

 Consider two projections $P$ and $Q$ satisfying $PQP=\lambda P$ with $0<\lambda<1$. As $0<\lambda<1$, $P$ and $Q$ cannot have a common eigenstate corresponding to value $(1,\mu)$ with $\mu=0,1$ and let $\Pi_{\mu}$ be projection to the subspace in which $P$ and $Q$ have common eigenvalue $(0,\mu=0,1)$.  By selecting a  basis under which $P$ is diagonal, we have the following block form 
\begin{equation}
 Q = \left( \begin{array}{cc} 
 PQP & PQ\bar{P} \\
 \bar{P}QP & \bar{P}Q\bar{P}
    \end{array} \right)\oplus \Pi_{1}
 = \left( \begin{array}{cc} \lambda I & C \\
 C^{\dag} & D\end{array} \right)\oplus \Pi_{1}
\end{equation}
for the other projection $Q$, where we have denoted $\bar{P}=I-P-\Pi_{0}-\Pi_1$ the projection to the subspace where $P,Q$ do not have common eigenstates. As $Q$ is also a projection, we have $Q^2=Q$, i.e.,
\begin{align}\label{pol1}
\left\{\begin{aligned}
\lambda &= \lambda^2 + CC^{\dag} \\
D &= C^{\dag}C+D^2\\
C &= \lambda C+CD 
\end{aligned}\right. 
\Rightarrow
\left\{\begin{aligned}
 CC^{\dag} &= \lambda(1-\lambda) \\
 C^{\dag}C &= D(1-D) \\
 CD &= (1-\lambda)C
    \end{aligned}\right.,
\end{align}
where we have omitted the identity matrix for simplicity.  By multiplying $C^\dagger$ from left of the equatoin $CD = (1-\lambda)C$, we have $0=D(1-D)(D-1+\lambda)$ so that the eigenvalues of $D$ are $0, 1, 1-\lambda$. Consider now a basis $\{|d_{i|0}\rangle, |d_{j|1}\rangle, |d_{k|\bar\lambda}\rangle\}$ of the complement of $P$, i.e., $\bar P$ under which $D$ is diagonal. All the elements of the row and column in which the eigenvalue $0$ is located must be $0$ since $Q$ is positive semidefinite.  If there is a non-zero element in the row, then there must be a complex conjugate of this non-zero element in the corresponding position in the column. In this subspace, the determinant will be less than zero. Also, all the row and column elements in which the eigenvalue $1$ is located must be $0$ since $Q^2=Q$. That is, $C|d_{j|\mu}\rangle=0$ so that $Q|d_{j|\mu}\rangle=\mu |d_{j|\mu}\rangle $ for $\mu=0,1$ which contradicts the definition of $\bar P$ in which $P$ and $Q$ have no common eigenstates. 
As a result, $D=1-\lambda$. Since $CC^\dagger$ and $C^\dagger C$ have the same number of nonzero eigenvalues, we obtain that $C$ is a square matrix, i.e., $P$ and $\bar P$ are of the same dimension $R=\tr P$. Moreover, by using unitary transformation $U=C/\sqrt{\lambda(1-\lambda)}$, we can choose a basis in $\bar P$ such that $C$ becomes diagonal with a single entry $\sqrt{\lambda(1-\lambda)}$ and
$Q=(Q_\lambda\otimes I_R)\oplus \Pi_1$ while in the same basis we have $P=P_0\otimes I_R$ where 
$$Q_{\lambda} = 
\begin{pmatrix}
    \lambda & \sqrt{\lambda(1-\lambda)} \\
 \sqrt{\lambda(1-\lambda)} & 1-\lambda \\
\end{pmatrix},\quad 
P_{0} = \begin{pmatrix}
1 & 0 \\
0 & 0 \\
\end{pmatrix}.$$ 
That is, two projections $P$ and $Q$ can be simultaneously block-diagonalized into 
some two-dimensional subspaces in addition to common eigenspaces with eigenvalues $(0,0)$ and $(0,1)$. As a result, all possible values of ordered pair $(\langle P\rangle_\rho,\langle Q\rangle_\rho)$ is the convex hull of the region of the ordered pair of values attained in each subspace.

In the two-dimensional subspace, we have effective two projections $Q_\lambda$ and $P_0$ and ranging over all possible qubit states, we have the following FiGUR
\begin{equation}
 \langle (2P_0-1)\cos\theta +(2Q_\lambda-1)\sin\theta \rangle_\rho\le 
 \sqrt{1+(2\lambda-1)\sin2\theta}
\end{equation}
 for aribtrary $\theta$. By differentiating over $\theta$, we obtain the attainable region of the ordered pair $(\langle P_0\rangle,\langle Q_\lambda\rangle)$ as
\begin{equation}\label{aa1}
 \langle 2P_0-1\rangle^2+\langle 2Q_\lambda-1\rangle^2+(2\lambda-1)^2-2(2 \lambda -1)\langle 2P_0-1\rangle\langle 2Q_\lambda-1\rangle\le 1
\end{equation}

Taking into account the values $(0,0)$ and $(0,1)$ for the ordered pair $(\langle P_0\rangle,\langle Q_\lambda\rangle)$ attained in the corresponding common eigenspace, we obtain the region of all possible values of ordered pair $(\langle P\rangle,\langle Q\rangle)$ as the convex hull of the ellipse defined by Eq. (\ref{aa1}) and two isolated points $(0,0)$ and $(0,1)$ as shown in Fig.\ref{fig:figur}. As a result we obtain 
$$\langle Q\rangle\le f_\lambda(\langle P\rangle), \quad f_\lambda(x)=\left\{\begin{array}{ll}1-(\sqrt{(1-\lambda)x}-\sqrt{\lambda(1-x)})^2&x\ge \lambda\\
1&x\le \lambda\end{array}\right..$$
Note that $\sqrt{1-f_\lambda(x)}=\sqrt{(1-\lambda)x}-\sqrt{\lambda(1-x)}$ for $x\ge \lambda$ we have the form of lemma in the main text.
Furthermore, if we have $QPQ=\lambda Q$, the local point $(0,1)$ is no longer attainable. The region is the convex hull of the ellipse defined by Eq.(\ref{aa1}) and point $(0,0)$, which is 
\begin{equation}
 \langle P\rangle+\langle Q\rangle-2\sqrt{\lambda \langle P\rangle\langle Q\rangle}\le 1-\lambda
\end{equation}

For a unitary operator satisfying $g^d=1$ we denote by
$\lceil g\rceil=\frac1d\sum_{k=0}^{d-1} g^k$
the eigenspace corresponding to eigenvalue 1 and $\ty g_\rho=\tr\rho\ty g$ its average in the state $\rho$. For two projections corresponding to +1 eigenspaces of two noncommuting generalized Pauli operators $g,h$ satisfying $gh=\omega^\eta hg$ with $\eta\not=0$ we have
$$\ty g\ty h\ty g=\frac1{d^2}\sum_{jk}g^jh^k\ty g=\frac1{d^2}\sum_{jk}h^k\omega^{\eta jk}\ty g=\frac1d\ty g$$
with $\lambda=1/d$. Simplarly we also have $\ty h\ty g\ty h=\ty h/d$. We thus have for our purpose:

{\bf Lemma 1'} {\it For two noncommuting strings of generalized Pauli operators $g,h$ it holds the following FiGUR
\begin{equation}
    \ty g_\rho \le f_{\frac1d}(\ty h_\rho) ,\quad \ty h_\rho\le f_{\frac1d}(\ty g_\rho) 
\end{equation}
where 
\begin{equation}
f_{\frac1d}(x) = \left\{\begin{array}{lc}1-\frac1d\left(\sqrt{{1-x}}-\sqrt{{(d-1)x}}\right)^2& x\ge \frac1d\\
1&x\le \frac1d\end{array}\right.
\end{equation}
is a decreasing function of $x$.}

\begin{figure}
    \includegraphics[height=5cm]{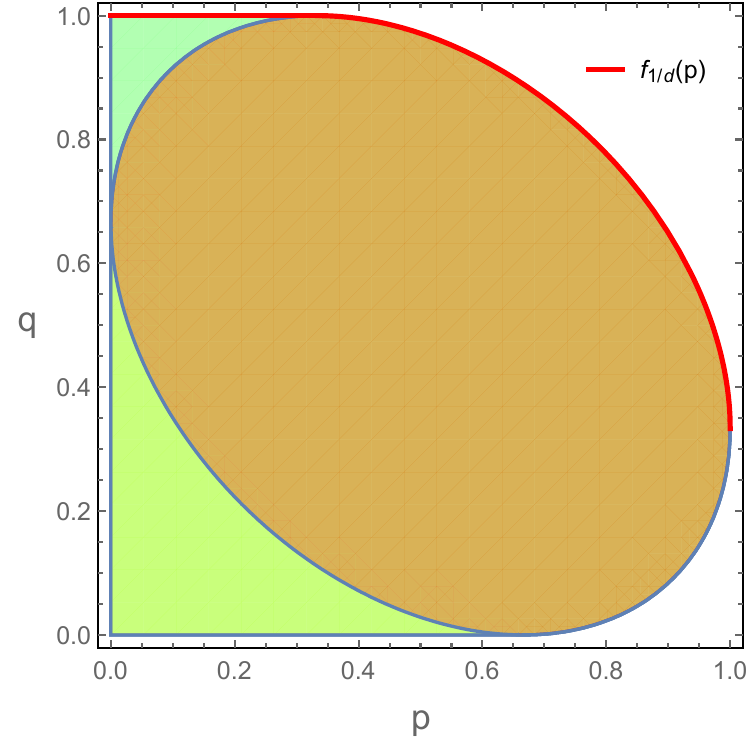}
    \caption{Fine-grained uncertainty relations for two projections and the corresponding decreasing function $f_{\lambda}(p).$}
    \label{fig:figur}
\end{figure}

{\bf Lemma 2.} {\it 
For two commuting $g,h$ it holds
\begin{equation}\label{2d5}
    \ty {gh}\ge \ty{g}\ty{h}\ge \ty g+\ty{h}-1
\end{equation}}

Proof. For the first inequality, we have
\begin{equation}
    \begin{aligned}
\ty{gh}&=\frac 1d\sum_k (hg)^k =\frac 1{d^2}\sum_{k,k',l}h^kg^{k'}\omega^{(k-k')l} \\
&=\sum_{l=0}^{d-1}\ty{\omega^l h}\ty{\omega^{-l}g} \ge \ty h\ty g \nonumber
    \end{aligned}
\end{equation}
The second inequality follows from the fact that $(1-\ty g)(1-\ty{h})\ge 0$ for two commuting projections.

\section{Proof of Theorem 2}
{\it Inflation.--- }A powerful tool to study LOSR networks is the quantum inflation technique \cite{wolfe2019inflation,navascues2020inflation,wolfe2021quantum}. An inflation of a given network is a network that consists of multiple copies of the original network whose sources might be shared by nodes from different copies. For our purpose here, we consider a LOSR network with bipartite sources shared between each pair of nodes labeled with $V$ and its inflations with only two copies $(V,V')$. Fixing a node, e.g., node number 2, and a subset $T\subset V$ of nodes, we redistribute the sources across two copies according to the following: each node $t\in T$ shares a bilocal source with node $2'\in V'$ instead of $2\in V$ and each node in $T'\subset V'$ shares a bipartite source with node number $2\in V$, while keeping all remaining nodes in the same copy fully sharing bipartite sources with each other, as shown in Fig.\ref{fig:inf}. In this case, an inflation can be specified by the subset $T$ alone.

\begin{figure}[h]
\includegraphics[width=0.45\textwidth]{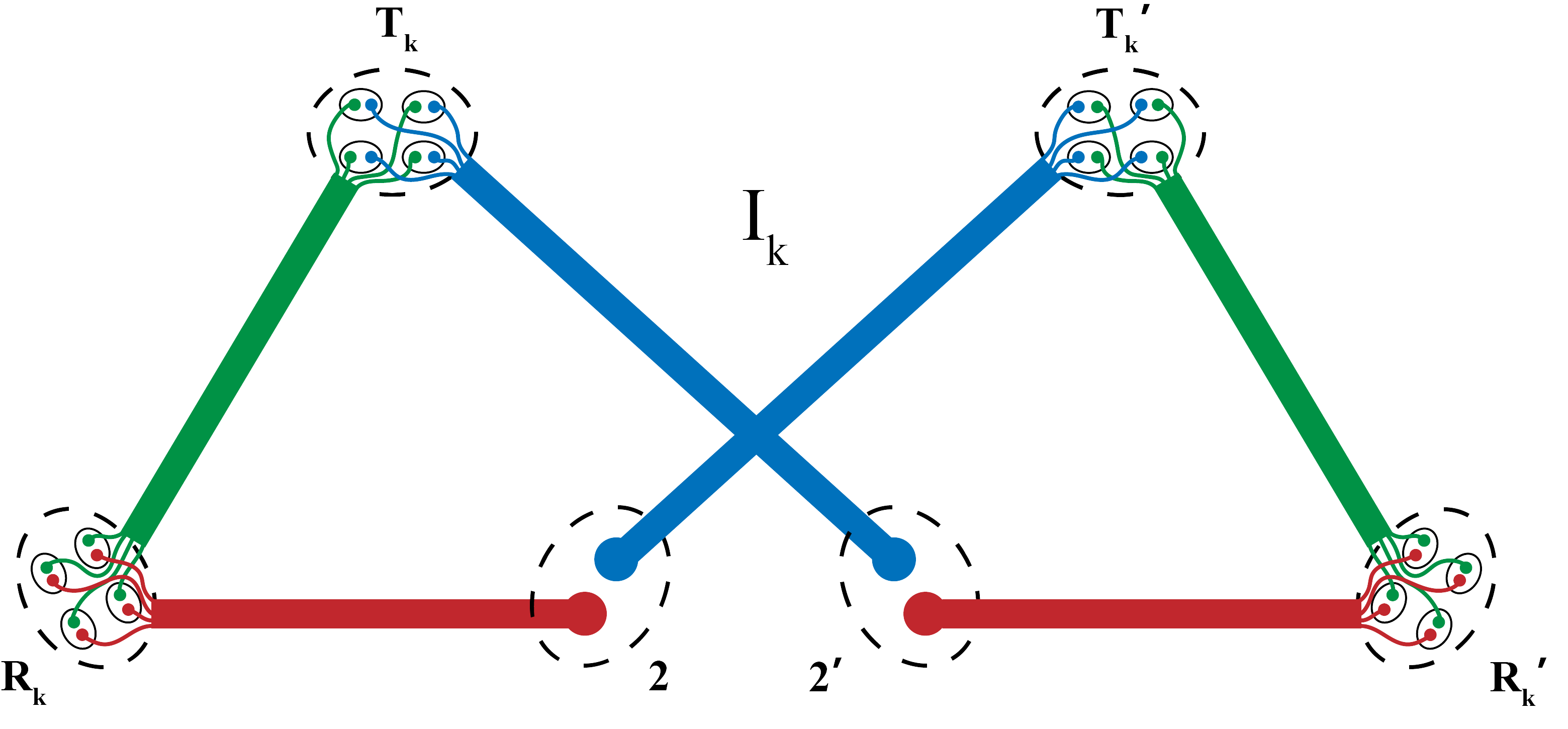}
\caption{An inflation $\mathcal{I}_k$ of a given quantum network on nodes $V=T_k\cup R_k\cup\{2\}$ specified by a subset $T_k\subset V$. Each pair of nodes within dashed ellipses shares a bipartite source.}
\label{fig:inf}
\end{figure}

{\it Proof.---}
Consider a graph in its standard form with index $\beta=2|N_1\cap N_2|+1$ where vertices 1 and 2 are neighbors. As ensured by Theorem 1, for each vertex $u\in \Delta_0=N_1\cap N_2$ there exists a vertex $j_u\in N_u$ disconnected from 2. Different $u$ might have the same extra neighbor $j_u$. We can then bound 
\begin{eqnarray*}
f_{\frac1d}(2F-1)&\ge&f_{\frac1d}(\ty{g_3^x}_\rho+\ty{g_3^{-x}g_2}_\rho-1)\\
\mbox{\small 1st }&=&f_{\frac1d}(\ty{g_3^x }_{T}+\ty{g_3^{-x}g_2}_{T}-1)\\
\mbox{\small 2nd }&\ge& f_{\frac1d}(\ty{g_2}_{T})\\
\mbox{\small 2nd }&=&
f_{\frac1d}(\ty{g_2}_{T_0})\\
\mbox{\small 3rd }&\ge& \ty{\tilde g_1}_{T_0}\\
\mbox{\small 3rd }&=& \ty{g_1}_{T_{1}}\\
\mbox{\small 4th }&\ge& \ty{g_{j_u}^{-x_u}}_{T_{1}}+\ty{g_{j_u}^{x_u}g_1}_{T_{1}}-1\\
\mbox{\small 4th }&=&\ty{g_{j_u}^{-x_u}}_\rho+\ty{g_{j_u}^{x_u}g_1}_{T_u}-1\\
\mbox{\small 5th }&\ge&\ty{g_{j_u}}_\rho+\ty{g_{j_u}}_{T_u}+\ty{g_1}_{T_u}-2\\
\mbox{\small 5th }&=&2(\ty{g_{j_u}}_\rho-1)+\ty{g_1}_{T_u}\\
\mbox{\small * }&\ge&\cdots \ge2\sum_{u\in \alpha\subset \Delta_0}(\ty{g_{j_u}}_\rho-1)+\ty{g_1}_{T_{\alpha}}\ge\cdots\ge2\sum_{u\in \Delta_0}(\ty{g_{j_u}}_\rho-1)+\ty{g_1}_\rho\\
&\ge&1-\beta(1-F)
\end{eqnarray*}
\begin{itemize}
\item The first inequality is because function $f_{\frac1d}(x)$ is decreasing and the fidelity $F=\langle G|\rho|G\rangle\le \ty{g}_\rho$ for any stabilizer of the graph state $|G\rangle$. 
\item We consider the inflation defined by $T=\{1\}$. The support of $g_3$ does not contain $2$ while that of $g_3^{-x}g_2$ does not contain $1$ for $x=\Gamma_{12}/\Gamma_{13}$ so that both supports $|g_3|$ and $|g_3^{-x}g_2|$ are fully connected in the inflation $T$ leading to $\ty{g_3}_\rho=\ty{g_3}_T$ and $\ty{g_3^{-x}g_2}_\rho=\ty{g_3^{-x}g_2}_T$.
\begin{figure}[h]
\includegraphics[height=3.3cm]{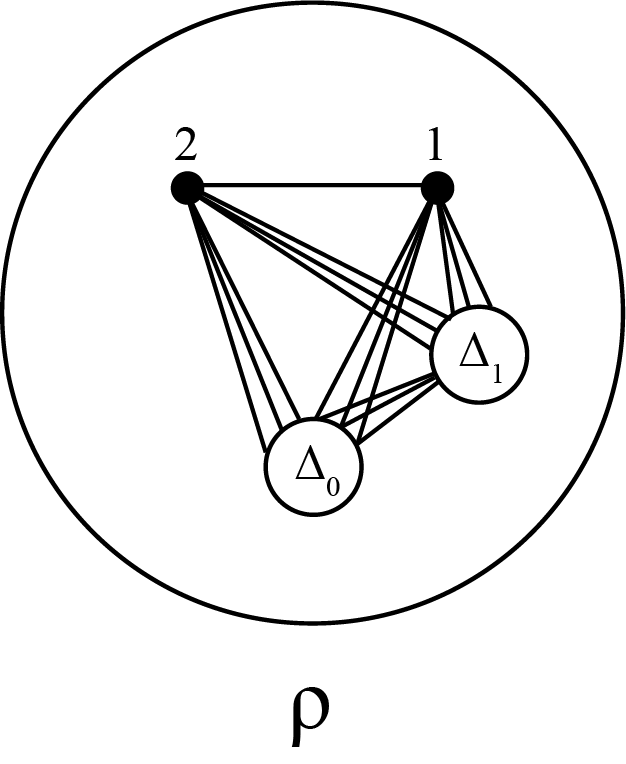}\hskip.5cm\includegraphics[width=0.33\textwidth]{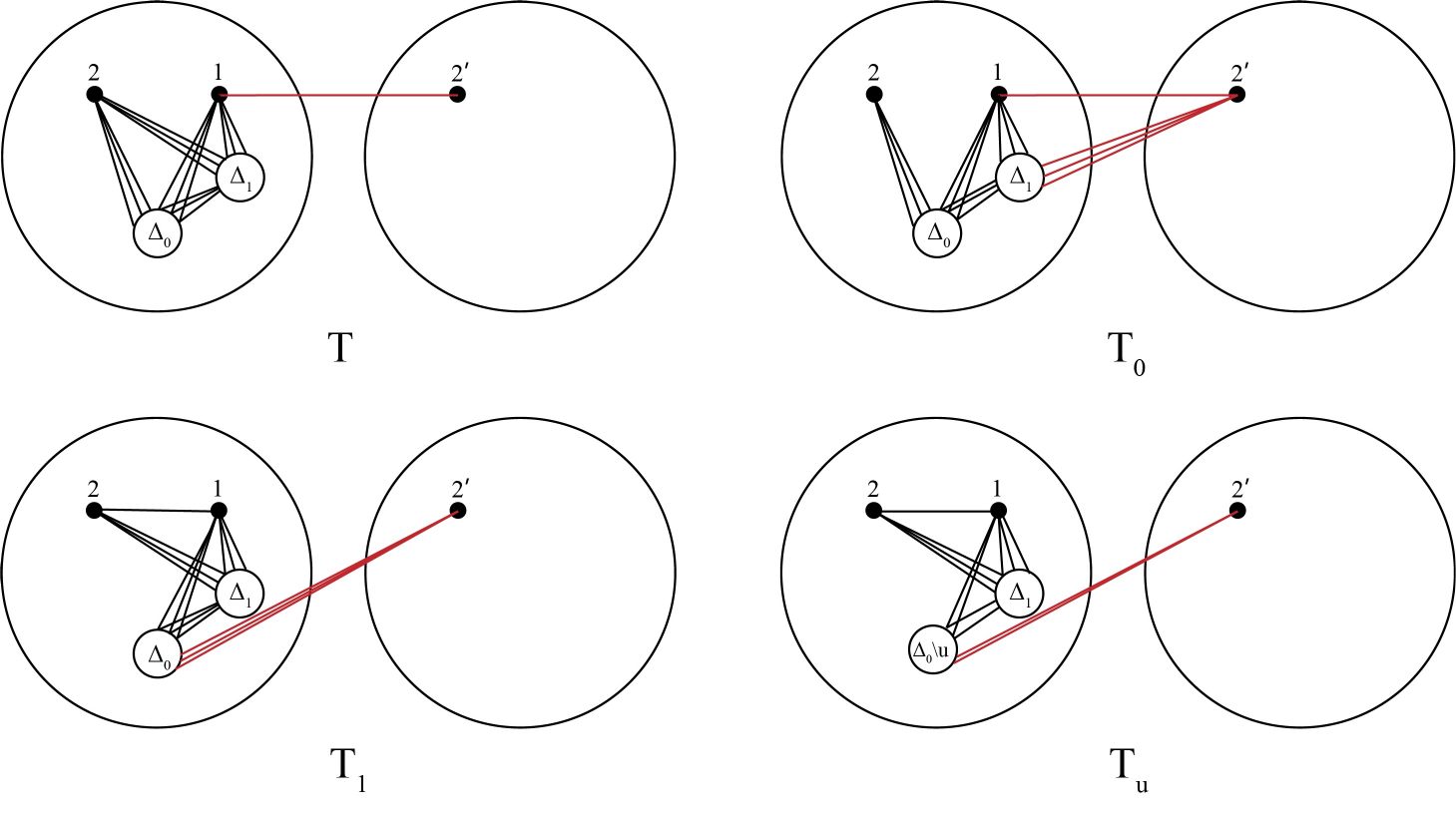}
\caption{The inflation generated by set $T=\{1\}$ in which vertex 1 shares a bipartite source with $2'$ instead of 2. Vertex 3 belongs to $\Delta_1$, the set of all neighbors of 1 different and disconnected from vertex 2.}
\end{figure}
\item The 2nd inequality is due to Lemma 2.
\item For the second equality we consider inflation $T_0=\{1\}\cup \Delta_1$ with $\Delta_1=N_1\backslash (\Delta_0\cup\{2\})$. As $|g_2|\cap \Delta_1=\emptyset$, the vertices in the support of stabilizer $g_2$ are fully connected except for vertices $1$ and $2$ in both networks $T_0$ and $T$ so that we have $\ty{g_2}_{T}=\ty{g_2}_{T_0}$.
\begin{figure}[h]
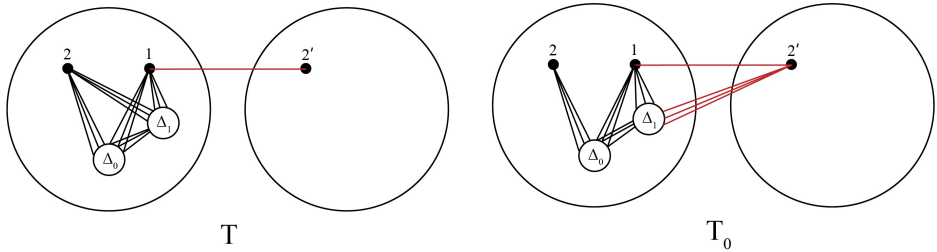

\includegraphics[width=0.33\textwidth]{infl1.pdf}\hskip.5cm\includegraphics[width=0.33\textwidth]{infl2.pdf}
\caption{The inflation generated by set $T_0=\{1\}\cup\Delta_1$ in which vertices in $\{1\}\cup\Delta_1$ share a bipartite source with $2'$ instead of 2. }
\end{figure}
\item The 3rd inequality is due to FiGUR applied to noncommuting generalized Pauli operators $g_2$ and $\tilde g_1$.

\item The 3rd equality is because vertices in the support $|\tilde g_1|=\{1,2'\}\cup\Delta_0\cup\Delta_1$ in inflation $T_0$  are fully connected except that vertex $2'$ is disconnected from its neighbors in $\Delta_0$ while vertices in the support $|g_1|=\{1,2\}\cup\Delta_0\cup\Delta_1$ in inflation $T_1=\Delta_0$ are fully connected except that vertex 2 is disconnected from its neighbors in $\Delta_0$.
\begin{figure}[h]
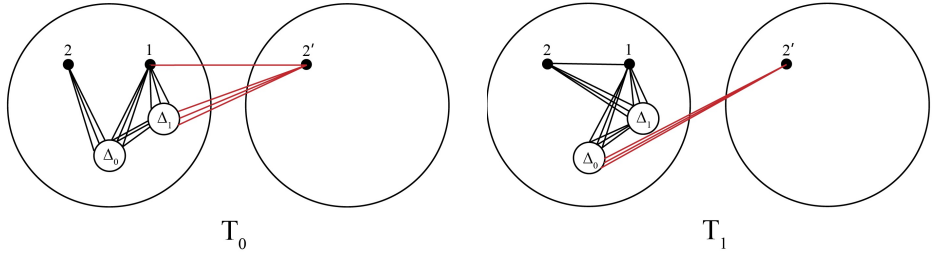

\includegraphics[width=0.33\textwidth]{infl2.pdf}\hskip.5cm\includegraphics[width=0.33\textwidth]{infl3.pdf}
\caption{The inflation generated by set $T_1=\Delta_0$ in which all vertices in $\Delta_0$ share a bipartite source with $2'$ instead of 2.}
\end{figure}
\item The 4th inequality is due to Lemma 2.
\item The 4th equality is because from $2\not\in|g_{j_u}|$ it follows that the support $|g_{j_u}|$ is fully connected so that $\ty{g_{j_u}}_{T_1}=\ty{g_{j_u}}$ while $u\not \in |g_{j_u}^{x_u}g_1|$ by choosing $x_u=-\Gamma_{1u}/\Gamma_{j_u,u}$ so that in both inflations $T_1$ and $T_u=\Delta_0\backslash \{u\}$ vertices in the support $|g_{j_u}^{x_u}g_1| $ share the same configuration of bipartite sources, leading to $$\ty{g_{j_u}^{x_u}g_1}_{T_1}=\ty{g_{j_u}^{x_u}g_1}_{T_u}.$$
\begin{figure}[h]
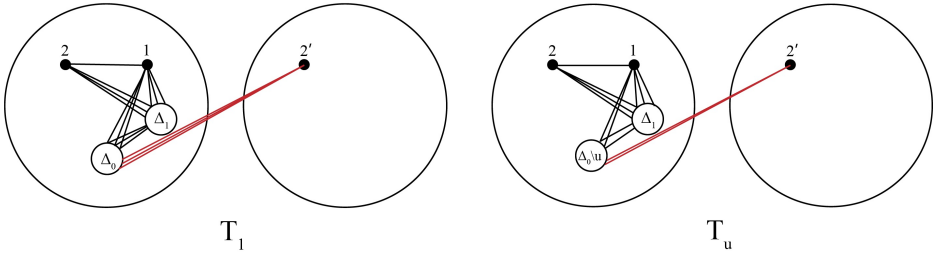

\includegraphics[width=0.33\textwidth]{infl3.pdf}\hskip.5cm
\includegraphics[width=0.33\textwidth]{infl4.pdf}
\caption{The inflation generated by set $T_u=\Delta_0\backslash \{u\}$ in which vertices in $\Delta_0\backslash \{u\}$ share a bipartite source with $2'$ instead of 2. }
\end{figure}
\item The 5th inequality is due to Lemma 2.
\item[*] We proceed by taking all vertices in $\Delta_0=\{u_1=u,u_2,\ldots,u_q\}$ one by one with $q=|\Delta_0|$ and considering the following sequence of inflations:
$$T_1=\Delta_0\supset T_2=\Delta_0\backslash \{u_1\}=T_u\supset T_3=\Delta_0\backslash \{u_1,u_2\}\supset\ldots\supset T_{q+1}=\emptyset.$$ We also denote $T_\alpha=\Delta_0\backslash \alpha$. By repeating the arguments for the 4th equality above in the case of $T_1\supset T_u=T_2$ for the cases $T_n\supset T_{n+1}$ with $n=2,3,\ldots,q$ we obtain
$$\ty{g_{j_n}}_{T_n}=\ty{g_{j_n}}_{T_{n+1}}=\ty{g_{j_n}}_{\rho},\quad \ty{g_{j_n}^{x_n}g_1}_{T_n}=\ty{g_{j_n}^{x_n}g_1}_{T_{n+1}}\ge\ty{g_{j_n}^{x_n}}_{T_{n+1}}+\ty{g_1}_{T_{n+1}}-1,$$
where $j_n$ is a neighbor of $u_n$ disconnected from 2 and $x_n=-\Gamma_{1,u_n}/\Gamma_{u_n,j_n}$.
\item Last inequality is due to $F\le \ty{g}_\rho$ and note that $\beta=2q+1$.
\end{itemize}

To arrive at the fidelity upper bound in Theorem 2, we note that
$$\beta (1-F)\ge 1-f_{\frac1d}(2F-1)=\frac1d\left(\sqrt{(d-1)(2F-1)}-\sqrt{2(1-F)}\right)^2$$
from which it follows whenever $F\ge\frac{d+1}{2d}$
$$\sqrt{d\beta (1-F)}\ge \sqrt{(d-1)(2F-1)}-\sqrt{2(1-F)}.$$




\textit{Genuine tripartite Nonlocality arises from network states.---} Here we consider the nonlocality and genuine tripartite nonlocality obtained with the network states from Protocol I with sources distributing $d$-dimensional particles, where $d=2, 3, 4$, the results are listed in Table~\ref{tab:nonlocality}.The two inequalities $g_1$ and $g_2$ to detect genuine tripartite nonlocality are as follows:
\begin{align}
    &2\langle C_1\rangle +2\langle B_1\rangle +2\langle B_1C_2\rangle +\langle A_1\rangle +\langle A_1C_1\rangle +\langle A_1B_1\rangle -2\langle A_1B_1C_1\rangle -\langle A_1B_1C_2\rangle +\langle A_1B_2C_1\rangle -\langle A_1B_2C_2\rangle\nonumber\\ 
    &+\langle A_2\rangle +\langle A_2C_1\rangle +\langle A_2B_1\rangle -2\langle A_2B_1C_1\rangle -\langle A_2B_1C_2\rangle -\langle A_2B_2C_1\rangle +\langle A_2B_2C_2\rangle \le 6
\end{align}
\begin{align}
    &2\langle C_1\rangle +\langle B_1\rangle +\langle B_1C_1\rangle +\langle B_2\rangle +\langle B_2C_1\rangle +\langle A_1\rangle +\langle A_1C_1\rangle +\langle A_1B_1\rangle -2\langle A_1B_1C_1\rangle +\langle A_1B_1C_2\rangle -\langle A_1B_2C_1\rangle\nonumber\\ 
    &-\langle A_1B_2C_2\rangle +\langle A_2\rangle +\langle A_2C_1\rangle -\langle A_2B_1C_1\rangle -\langle A_2B_1C_2\rangle +\langle A_2B_2\rangle -2\langle A_2B_2C_1\rangle +\langle A_2B_2C_2\rangle \le 6
\end{align}
\begin{table}[h]
  \centering
\begin{tabular}{lccccc}
\hline\hline
 \# & \text{C} & \text{Q} & $d=2$ & $d=3$ & $d=4$ \\ \hline 
 4 & 2 & 3.65685 & 2.00211 & 1.99962 & 1.98873 \\
 5 & 3 & 4.88854 & 3.00905 & 3.02612 & 3.01511 \\
 6 & 3 & 4.65685 & 3.00411 & 3.00752 & 2.99420 \\
 21 & 4 & 5.95546 & 4.00545 & 4.01432 & 4.00016 \\
 40 & 6 & 8.12979 & 6.00715 & 6.01344 & 5.98919 \\
{$g_1$} & 6 & 6.82507 & 6.00001 & 5.97618 & 5.93736 \\
 {$g_2$} & 6 & 6.56259 & 6.00058 & 5.97618 & 5.93736 \\  \hline \hline 
\end{tabular}
\caption{The violation of tripartite Bell inequalities with triangle network states prepared according to protocol I. Here the number $i$ in the first column stand for Sliwa's $i$-th inequality~\cite{sliwa2003symmetries}. The notations $g_1$ and $g_2$ stand for two genuine tripartite Bell inequalities as found in Ref.~\cite{bancal2013definitions}. The columns labeled by C and Q list the classical and quantum optimal values of those inequalities. The columns with $d=2$ corresponds to the optimal value with the triangle network state prepared with $2$-dimensional entangled pairs as sources according to protocol I, similarly for the columns labeled by $d=3$ and $d=4$.}\label{tab:nonlocality}
\end{table}

\subsection*{$d=2$ protocol}

\begin{itemize}

\item In general the measurement reads
$${\Pi_{ab}=\sum_{\sigma=0}^{1}|a+\sigma,b+\sigma\rangle\langle a+\sigma,b+\sigma|,\quad (a,b)\in M:=\{(a,b)\mid \min\{a,b\}\equiv 0\mod 2\}}$$
  with the convention $|j\rangle=0$ for $j\ge t$. It is complete
  $$\sum_{(a,b)\in M}\Pi_{ab}=I_{t^2}.$$
  \item  In the case of source dimension $t=6$ and target dimension $d=2$, the measurements are shown as $1,2$-dimensional projections indicated by groupped dots
\begin{center}
\includegraphics[width=0.2\textwidth]{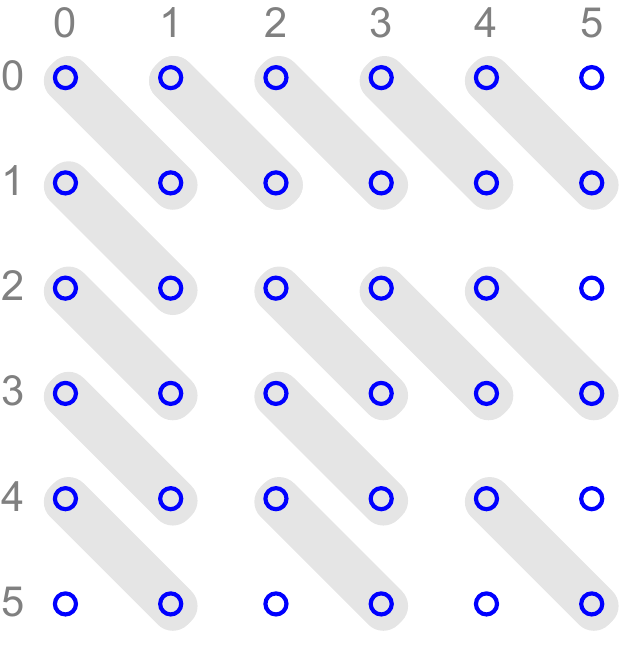}
\end{center}
\item For each node the encoding reads
$$E_{ab}=\sum_{\sigma=0}^{1}|\sigma\rangle\langle a+\sigma,b+\sigma|,\quad (a,b)\in M$$
which defines a trace-preserving isometry
$$\mV:\mH_{t^2}\mapsto \mH_2,\quad \mV(\rho)=\sum_{(a,b)\in M}E_{ab}\rho E_{ab}^\dagger$$
\item With bipartite sources of local dimension $t$
$$|\gamma\rangle_{12'}=\sum_{j_3=0}^{t-1}\gamma_{j_3}|j_3j_3\rangle_{12'},\quad |\alpha\rangle_{23'}=\sum_{j_1=0}^{t-1}\alpha_{j_1}|j_1j_1\rangle_{23'},\quad |\beta\rangle_{31'}=\sum_{j_2=0}^{t-1}\beta_{j_2}|j_2j_2\rangle_{31'}$$
a pure network state reads
$$|\psi\rangle=|\gamma\rangle_{12'}\otimes|\alpha\rangle_{23'}\otimes|\beta\rangle_{31'}
=\sum_{j_1,j_2,j_3}\alpha_{j_1}\beta_{j_2}\gamma_{j_3}|j_3j_2\rangle_{11'}\otimes|j_1j_3\rangle_{22'}\otimes|j_2j_1\rangle_{33'}$$
\item Its fidelity with the 3-qubit GHZ state reads 
\begin{eqnarray*}
F_{t}&=&\langle GHZ|\mV_1\otimes \mV_2\otimes \mV_3(|\psi\rangle\langle\psi|)|GHZ\rangle\\
&=&\sum_{(a_1,b_1),(a_2,b_2),(a_3,b_3)\in M}\langle GHZ|E_{a_1b_1}\otimes E_{a_2b_2}\otimes E_{a_3b_3}|\psi\rangle\langle\psi|E_{a_1a_1}^\dagger\otimes E_{a_2b_2}^\dagger\otimes E_{a_3b_3}^\dagger|GHZ\rangle\\
&=&\frac12\sum_{(a,b),(b,c),(c,a)\in M}\left(\textstyle\sum_{\sigma=0}^{1}\alpha_{a+\sigma}\beta_{b+\sigma}\gamma_{c+\sigma}\right)^2
\end{eqnarray*}
with the convention $\alpha_j,\beta_j,\gamma_j=0$ whenever $j\ge t$.
\item Maximal fidelity $F^*_{t}=\max_{|\psi\rangle}F_{t}$ can readily found by numerical method for $2\le t\le 12$
$$\{0.51704,0.540053,0.545959,0.547493,0.5479,0.548009,0.548038,0.548045,0.548047,0.548048,0.548048\}$$

\end{itemize}

\begin{figure}
  \includegraphics[width=0.45\textwidth]{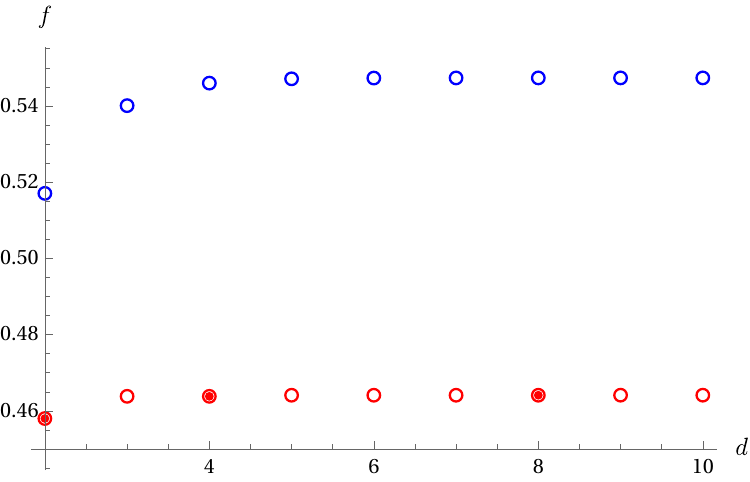}
  \caption{The fidelity of the network states prepared according to Protocol I and Protocol II, where $d$ stands for dimension and $f$ represents fidelity, the blue circles correspond to $2$-dimensional GHZ state, the red circles and red points correspond to $3$-dimensional GHZ state with general states and separable pairs, respectively.}
  \label{fig:fidelity}
\end{figure}

\newpage

\bibliographystyle{apsrev4-2}
\bibliography{references}
\end{document}